%% file: main.tex
\documentclass[sigplan,10pt]{acmart}

\renewcommand\footnotetextcopyrightpermission[1]{}
\usepackage{tikz}
\usepackage{amsmath}
\usepackage{paralist}

\input{custom.tex}

\begin{document}

\date{}

\title{Nyx-Net: Network Fuzzing with \\ Incremental Snapshots}
\author{
{Sergej Schumilo$^1$, Cornelius Aschermann$^1$, Andrea Jemmett$^2$, Ali Abbasi$^1$ and Thorsten Holz$^1$}\\
$^1$Ruhr-Universit\"at Bochum\\
$^2$Vrije Universiteit Amsterdam\\
}

\settopmatter{printfolios=true}

\input{sections/abstract}

\maketitle

\input{sections/introduction}
\input{sections/background}

\input{sections/design}
\input{sections/implementation}
\input{sections/evaluation}
\input{sections/related_work}
\input{sections/conclusion}


\input{main.bbl}
\input{sections/appendix}

\end{document}

%% file: custom.tex
\PassOptionsToPackage{hyphens}{url}
\usepackage{tcolorbox}
\usepackage{graphicx}
\usepackage[font=footnotesize]{caption}
\usepackage{color, colortbl}
\usepackage{subcaption}
\usepackage{listings}
\usepackage{amsmath, amsthm, amsfonts}
\usepackage{tikz}
\usepackage[linesnumbered,ruled,vlined]{algorithm2e}
\usepackage{csquotes}
\usepackage{multirow}
\usepackage{multicol}
\usepackage{tabularx}
\usepackage{booktabs}
\usepackage[flushleft]{threeparttable}
\usepackage[frozencache=true,cachedir=.]{minted}

\usepackage{pifont}
\usepackage{tablefootnote}
\usepackage{tikz}

\usepackage{adjustbox}

\SetAlFnt{\footnotesize}

\input{commands}
\usepackage[hyphens]{url}

%% file: commands.tex
\newcommand{\tool}{\textsc{Nyx-Net}\xspace}
\newcommand{\eg}{{e.g.,}\xspace}
\newcommand{\libc}{{\tt libc}\xspace}
\newcommand{\agamotto}{\textsc{Agamotto}\xspace}
\newcommand{\firefox}{Firefox\xspace}
\newcommand{\afl}{\textsc{AFL}\xspace}
\newcommand{\aflsmart}{\textsc{AFLsmart}\xspace}
\newcommand{\aflnet}{\textsc{AFLnet}\xspace}
\newcommand{\aflnwe}{\textsc{AFLnwe}\xspace}
\newcommand{\aflpp}{\textsc{AFL++}\xspace}
\newcommand{\libpreeny}{\textsc{libpreeny}\xspace}
\newcommand{\ijon}{\textsc{Ijon}\xspace}
\newcommand{\syzkaller}{\textsc{syzkaller}\xspace}
\newcommand{\vdf}{\textsc{VDF}\xspace}
\newcommand{\profuzzbench}{\textsc{ProFuzzBench}\xspace}

\newcommand{\libfuzzer}{\textsc{libFuzzer}\xspace}
\newcommand{\peach}{\textsc{Peach}\xspace}
\newcommand{\kafl}{\textsc{kAFL}\xspace}
\newcommand{\nyx}{\textsc{Nyx}\xspace}

\newcommand{\nautilus}{\textsc{Nautilus}\xspace}
\newcommand{\kvm}{{\textsc{KVM}}\xspace}
\newcommand{\qemu}{{\textsc{QEMU}}\xspace}

\newcommand{\cmark}{\ding{51}}

\lstMakeShortInline[columns=fixed]|

\newcommand{\pafig}[3]{ 
\begin{figure}[t]
  \centering
      \includegraphics[width=0.48\textwidth]{#1}
  \caption{#3}
  \label{#2}
  \vspace{-0.4cm}
\end{figure}
} 

\newcommand{\pafigwide}[3]{ 
\begin{figure*}[t]
  \centering
      \includegraphics[width=0.93\textwidth]{#1}
  \caption{#3}
  \label{#2}
  \vspace{-0.4cm}
\end{figure*}
}

\definecolor{trolleygrey}{rgb}{0.38, 0.38, 0.38}

\lstdefinestyle{Clang}{language=C,
    basicstyle=\scriptsize\ttfamily,
    commentstyle=\color{trolleygrey}\ttfamily,
    rulecolor=\color{Black},
    numberstyle=\tiny\color{Black},
    stepnumber=1,
    emph={int, u64, INT, char, double, float, unsigned, void, VOID, MSG, CALLBACK},
    emphstyle={\color{VioletRed}},
    stringstyle={\color{OliveGreen}},
    numbersep=8pt,
    showstringspaces=false,
    breaklines=true,
    frameround=ffff,
    frame=single,
        tabsize=4,
        captionpos=b
}

\usepackage{graphicx}

%% file: sections/abstract.tex
\begin{abstract}

Coverage-guided fuzz testing (``fuzzing'') has become mainstream and we have observed lots of progress in this research area recently. However, it is still challenging to efficiently test network services with existing coverage-guided fuzzing methods. In this paper, we introduce the design and implementation of \tool, a novel snapshot-based fuzzing approach that can successfully fuzz a wide range of targets spanning servers, clients, games, and even \firefox's Inter-Process Communication (IPC) interface.
Compared to state-of-the-art methods, \tool improves test throughput by up to 300x and coverage found by up to 70\%. Additionally, \tool is able to find crashes in two of \profuzzbench's targets that no other fuzzer found previously. When using \tool to play the game \emph{Super Mario}, \tool shows speedups of 10-30x compared to existing work. Under some circumstances, \tool is even able play ``faster than light'': solving the level takes less wall-clock time than playing the level perfectly even once. \tool is able to find previously unknown bugs in servers such as Lighttpd, clients such as MySQL client, and even \firefox's IPC mechanism---demonstrating the strength and versatility of the proposed approach. 
Lastly, our prototype implementation was awarded a \$20.000 bug bounty for enabling fuzzing on previously unfuzzable code in Firefox and solving a long-standing problem at Mozilla.
\end{abstract}

%% file: sections/introduction.tex
\section{Introduction}

In the last years, we have seen a lot of research progress in the field of fuzzing---both from academia~\cite{weizz, angora,qsym,tfuzz, drillerfuzzer, pangolin, ijon, nyx, stochfuzz,winnie} as well as industry~\cite{afl,aflpp,libfuzzer,lafintel}. 
The majority of improvements made in the last years have focused on improving fuzzing algorithms themselves. 
However, it is slowly becoming apparent that improvements for fuzzing on the algorithmic level have less impact in practice compared to improvements to the ability to fuzz new targets.
Any application that is fuzzed for the first time is likely to result in many security-relevant findings~\cite{exponentialcost}. 
Hence, we are currently observing a shift towards making it feasible (and even easy) to target new applications or systems. 

\begin{figure}[ht]
  \centering

\begin{tcolorbox}

\footnotesize

\emph{``Hence, we believe that a system testing approach is the only viable solution for [...] IPC testing.
One [...] approach [...] could be to [...] perform a snapshot of the parent [..] and then replace [..] child
messages [...]''}

--- \textbf{Blogpost} \footnotesize{- Security Team at Mozilla}

\end{tcolorbox}
    \caption[Excerpt taken from Mozilla's statement.]%
    {Excerpt taken from Mozilla's blogpost~\cite{mozilla-ipc} on IPC-Fuzzing.}
	\label{fig:mozilla}
\end{figure}

The two most common approaches to achieve this goal are generating function-style harnesses for persistent mode fuzzers such as \libfuzzer~\cite{libfuzzer,fuzzgen,fudge,winnie} or fuzzing based on snapshots taken at the start of the test case~\cite{nyx,agamotto,cmilk}. 
While these approaches address some of the problems of harnessing targets, none of them allows proper harnessing of systems with many messages that are being passed back and forth (e.g., network services). 
Such services pose a unique set of challenges distinct from ``parse a binary blob'' fuzzing targets: the target applications are typically much more stateful, more complex (and hence slower), and the message formats are often more complex than individual file formats.

\pafigwide{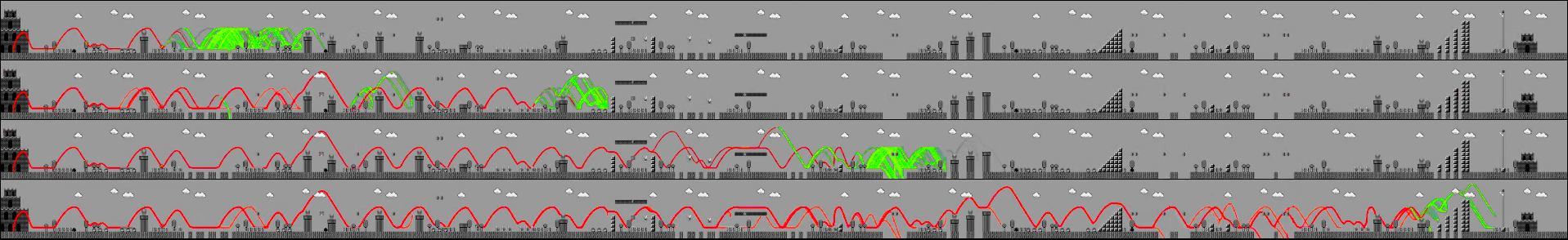}{fig:smbc}{Visualizing \tool' use of incremental snapshots to solve a hard Super Mario Bros. level: red paths are taken once to create an incremental snapshot. 
For each snapshot, a number of test cases (green paths) are performed. 
Note how \tool also uses ``old'' positions to create snapshots.}

Such applications make up a significant fraction of interesting attack surface. Yet, little has been done to allow effectively fuzzing and harnessing such targets with coverage-guided fuzzers. The notable exception is \aflnet~\cite{aflnet}, which runs the target and creates new connections for each test input. 
While this approach has long been used to fuzz network services in blind fuzzers, its drawbacks has made coverage-guided network fuzzing exceedingly difficult.
First, using network connections is significantly slower than reading from a file. 
Second, as the service is running persistently, the fuzzer has no clear point at which the service is ready to receive a test case; such fuzzers require waiting for manually-specified, fixed periods of time during startup, and use similar timeouts when handling each test case. 
Third, reusing the same process (without fully restarting the server) is also ``noisy'': for example, background threads in the service can randomly get scheduled independently of the test cases the fuzzer sent. 
These seemingly random code paths still affect the fuzzer's coverage and introduce pointless inputs into the queue.
Similarly, great care has to be taken to ensure that the target is not operating based on state introduced by previous test cases. 
To this end, \aflnet requires a user to write a ``cleanup'' script that is solely responsible for ensuring that all changes to the file system, databases, etc. are rolled back after each test case. 
Building such a script based on spurious, non-reproducible inputs can require a significant effort.

In this paper, we present the design and implementation of \tool, a fuzzing method that is able to test complex, stateful message-passing systems such as network services. Our approach is based on the following design principles: first, we use \emph{hypervisor-based snapshot fuzzing}~\cite{kafl, nyx, agamotto} to ensure noise-free fuzzing and to speed up resetting to a clean slate. Second, we propose \emph{selective emulation} of network functionality to avoid the heavy cost of real network traffic. This also allows \tool to control exactly when packets are consumed in order to avoid guessing the right timeouts. 

Our prototype implementation of \tool is built on top of \nyx~\cite{nyx}, which it enhances with a set of new capabilities. 
Most importantly, we enable \nyx to target network connections and add support for handling the network stack. We also extend \nyx's snapshot capabilities by introducing incremental, whole-VM snapshots. The snapshot mechanism is agnostic of the OS running in the VM. 
Using such snapshots increases performance and ensures that all state is properly reset between each test case. Furthermore, using such snapshots allows us to emulate a significant fraction of the network APIs: we run the target until a snapshot is taken without major interference (i.e., close to native speed). 
When the hooks detect that the target is about to receive the first bytes of fuzzer-supplied input, we take a whole-system snapshot. 
Following this snapshot, we now \emph{emulate} the network interactions of the target connection to further increase speed. 
As our evaluation shows, precisely emulating all network interactions is a difficult task. 
By only emulating the (few) operations on the target connection, we reduce the need to emulate all I/O functionality faithfully.
While we tested our prototype implementation only on two different Linux VM setups (\emph{busybox} and \emph{Ubuntu}) as well as targets running inside a Docker container inside of a Ubuntu VM, \tool's network emulation layer should be compatible with any POSIX-compliant system. 

Note that \tool is not limited to network interface fuzzing. Instead, our approach can fuzz any complex, stateful message-based target such as Inter-Process Communication (IPC) interfaces. For example, Firefox splits safety-critical parts of its codebase into isolated sandbox processes that communicate via various IPC methods such as Unix domain sockets and shared memory. We show that \tool can efficiently and effectively test these IPC interfaces.
In fact, after seeing that Mozilla was looking for a tool like \tool, we reached out to them and they decided to integrate \tool into their testing pipeline after finding multiple security issues using. An excerpt from Mozilla's blogpost on the challenges of IPC fuzzing~\cite{mozilla-ipc} is shown in Figure \ref{fig:mozilla}. 

As our evaluation shows, \tool drastically improves upon the state-of-the-art: compared to \aflnet on their own benchmark \profuzzbench~\cite{profuzzbench}, we are able to improve test throughput by up to 300x and coverage found by up to 70\%.
Compared to \agamotto~\cite{agamotto}, the state-of-the-art in snapshot fuzzing kernel modules, \tool is able to perform both snapshot reload and creation operations almost 10x faster.
Additionally, \tool is able to find crashes in two of \profuzzbench's targets that no other fuzzer could detect previously. 
In an evaluation with the game \emph{Super Mario Bros.}, we show that \tool is able to solve most levels about 10-30x faster than the state-of-the-art (\afl + \ijon~\cite{ijon}) --- demonstrating it's ability to improve the performance of message-based targets unrelated to networking (see Figure~\ref{fig:smbc} for a visualization).
In fact, when running in parallel, \tool is able to solve some levels ``faster than light'': solving the level takes less wall-clock time than playing the level perfectly even once. \tool is even able to exploit a glitch to solve a level that the authors of \ijon believed to be unsolvable. Lastly, \tool is not only able to find previously unknown bugs in servers such as Lighttpd, but also network clients such as MySQL client, and even \firefox's IPC. 

\smallskip \noindent 
In summary, we make the following key contributions:

\begin{compactitem}
\setlength\itemsep{0em}
\item We introduce \tool, an efficient fuzzing method that uses hypervisor-based snapshot fuzzing and selective emulation of network functionality to avoid the heavy cost of handling the full network traffic.

\item We study the concept of incremental snapshots in fuzzing for complex and slow targets, and show that our approach is able to efficiently test different types of client and server systems.

\item In our evaluation, we show that \tool outperforms state-of-the-art fuzzing tools by more than an order of magnitude in many benchmarks. 
Furthermore, our prototype implementation found multiple unknown bugs in complex, real-world software.

\end{compactitem}

\smallskip
To foster research, we will release \tool under an open-source license at \url{https://github.com/RUB-SysSec/nyx-net}.

%% file: sections/background.tex
\section{Technical Background}

We now discuss the technical difficulties of fuzzing real-world network services. We specifically focus on the approach used by existing methods to perform network fuzzing.
Given that \tool is based on \nyx, we also elaborate on the technical aspects of \nyx relevant for this paper.

\subsection{Network Service Fuzzing}

Many widely used network services and servers are still written in memory unsafe languages for performance reasons. This puts such software at significant risk: Most of the complexity is part of some very public (often Internet-wide) attack surface. To make the matter worse, getting memory safety right in languages that do not enforce memory safety is notoriously difficult.  
Over the last years, fuzzing has become one of the primary tools for finding complex memory safety bugs in an automated way. 
Of course, this includes network service fuzzing. For example, \aflnet~\cite{aflnet} removed \afl's focus on single files and enabled it to send packets to target sockets. There have also been various efforts to remap socket-based I/O to file-based I/O~\cite{libpreeny}. When successful, this approach allows to use pre-existing file-based fuzzers such as \aflpp~\cite{aflpp} to fuzz network services. 

However, effectively fuzzing network services remains challenging: 
\afl and most of its derivatives assume that the target is fast and spawns only a single process that runs until the input is consumed. 
Afterwards, the target is supposed to terminate immediately. 
Also, two sub-sequential executions should be (mostly) independent of each other. 
Unfortunately, none of these assumptions hold for most network services.
They are often designed with little regard for startup time, persist across connections (often by spawning threads or sub-processes), and maintain a significant state. 

\aflnet forces a user to write clean-up scripts that need to reset the environment to avoid contaminating results of later tests. It also employs fixed sleep times to ensure servers are online and in a good state. Overall, this makes it difficult to test new software and drastically reduces the performance. We found that it is not uncommon for \aflnet to only be able to achieve single digit test executions per second. 

To make matters worse, network APIs provided by current operating systems (OSs) are notoriously slow. 
Establishing a connection and reading data from it is far slower than reading from a file. 
A common workaround is to avoid network interfaces entirely: \libpreeny~\cite{libpreeny} introduced a ``de-sock'' hook that returns the file descriptor of \texttt{stdin} instead of network sockets when a new connection is established (this idea is now also part of \aflpp). 
This massively improves the performances, but fidelity is low: the vast majority of operations possible on sockets are not supported by \texttt{stdin}. 
As such, it will simply not work with most real-world software. 
Note that \libpreeny also contains a more advanced \texttt{stdin}-to-socket connection that uses a real socket and a new thread to move data from \texttt{stdin} to the network socket used by the target. 
This makes fuzzing of more complex network targets possible, but also losses the performance gains coming from network emulation. 

\subsection{Protocol Fuzzing}
Fuzzing network services is further complicated by the fact that they are much more interactive than software that processes static file formats.
Network services also often incorporate features such as compression, encryption, sequence numbers, and checksums that greatly hinder fuzzing efforts. 

Historically, this problem has been addressed by \emph{blind gene\-ra\-tor-style fuzzers}: the user simply writes a program that connects to the target and sends random, but (almost) valid protocol runs. 
This makes it easy to fix the aforementioned problems, but requires expert knowledge of the specific protocol and significant efforts. 

The main success criterion of \afl was that only a superficial or even no understanding of the format being fuzzed is necessary to use the tool to find bugs. 
Using \afl requires a set of start inputs (so-called \emph{seeds}). 
While \afl is often able to work even with empty seeds, it usually is more effective if sensible seeds are provided. 
This is mostly due to the mutation-based fuzzing of \afl and its coverage feedback. 
However, to use these advantages, in addition to writing a good generator, the user would also have to write a good mutator---an additional hurdle to jump that reduces usage of fuzzing. 
Nonetheless, there quite a few fuzzers exist that allow the user to specify detailed formats used for coverage-guided fuzzing. 
Usually, these fuzzers allow the user to provide a grammar or format specification~\cite{aflsmart,nautilus}. 

Commonly, the inputs all have to be valid for parsing. 
As a consequence, \aflsmart~\cite{aflsmart} (which uses \peach's~\cite{peach} pit file format to specify inputs) has to take great care to handle broken or otherwise unparseable inputs. Hence, \aflsmart only parses the seed inputs as it is computationally infeasible to parse new inputs found during fuzzing. 

Some fuzzers avoid this problem by simply making it impossible to use seed inputs. For example, \nautilus~\cite{nautilus} uses context-free grammars, but does not allow to provide seeds. 
As such, arbitrary grammars including those that are hard to parse can be used. 
Similarly, \nyx and \syzkaller~\cite{syzkaller} follow the purely generative approach and allow to specify input formats as sequences of typed function calls (or opcodes), but forgo the option to provide seeds. 

Lastly, in a far less principled, but just as effective approach, most fuzzers support registering custom mutators. 
They usually parse inputs on a best effort basis (\eg by splitting the input at newlines or matching parenthesis), perform some mutations, and then recreate the input. 
This unloads all the work to the user, but can be highly effective, particularly for formats where most structural information can be easily inferred (\eg line based formats). 
\aflnet follows a similar approach: it uses mutators based on a handful of rudimentary packet boundary parsers for the supported formats. 

\paragraph{\nyx's Affine Typed Bytecode}
As mentioned before, \nyx follows the generative approach: the user specifies a set of opcodes that can be chained by \nyx. 
While the authors only used the tool to fuzz hypervisors, the opcode-based approach can potentially be used to fuzz a wide variety of interactive targets. 
All the user needs to do is to implement a set of different opcodes (with their respective inputs and outputs). 

For example, a network specification handling multiple connections at the same time is shown in Listing~\ref{lst:netmulti}. 
First, we define a data type (\texttt{d\_bytes}) that contains the payload of actual packets. 
Then, we define a new opcode (or ``node'' in \nyx's terminology) that creates a new connection. 
It takes no inputs and returns a new connection handle (\texttt{e\_con}). 
Lastly, we define an opcode that emits/sends a single packet via a given connection. 
To this end, we create a node that borrows a connection and contains a vector of bytes. 

Note that the specification that we use for network targets in this paper is even simpler: we usually hook the first connection established via a given port and address. 
Our agent then delivers packets to each function call that attempts to read data from this connection (e.g. \texttt{recv()} or \texttt{read()}). 
Similarly, the agent signals readiness when functions such as \texttt{epoll()} or \texttt{select()} try to wait for more data on the given connection. 
All that remains is to fill out the two opcode handlers with actual C code that establishes a connection and sends the packet. 
The fuzzer auto-generates a bytecode format and a custom VM that executes the bytecode by calling the corresponding handlers, as well as custom mutators. 

\begin{listing}[t]
\begin{minted}{py}
d_bytes = s.data_vec("bytes", s.data_u8("u8"))
n_con = s.node_type("connection", outputs=[e_con])
n_pkt = s.node_type("pkt", borrows=[e_con], d_bytes)
\end{minted}
\caption{A (hypothetical) specification for multi-connection network emulation.}
\label{lst:netmulti}

\end{listing}

\subsection{Hypervisor-Based Snapshot Fuzzing}
As explained before, many network applications maintain state between individual test cases or have expensive startup routines. 
The former reduces reproducibility, while the latter reduces test throughput. 
It turns out that both problems can be largely avoided by a clever trick: by obtaining a snapshot of the system's state directly \emph{before} executing the test case, we can reset the system to a deterministic state after each test. 
The cost of this reset is independent of startup complexity and only determined by the size of the changes to the state of the system caused by executing the test.  
For example, while starting \firefox requires to load hundreds of megabyte of code into memory and to initialize all kinds of system APIs, handling a handful of IPC packets will typically only dirty a few hundred kilobytes of memory. 
\aflpp contains a Linux kernel module that is able to reset the memory and some limited kernel state of target ring-3 processes to increase performance. 
\agamotto\cite{agamotto} and \nyx both implement such a mechanism to create a snapshot of a whole VM and to reset back to this snapshot after each test. 
This allows efficient and deterministic fuzzing of a whole OS and even hypervisors. 
As \tool is based on \nyx, we now give a short introduction into how \nyx captures and reapplies VM snapshots. 

\paragraph{\nyx Agents}
\nyx runs the fuzz target inside of a VM controlled by a modified version of \qemu, and executed by a modified build of \kvm.
\qemu sets up the VM state and emulates devices as needed, while \kvm uses hardware virtualization extensions provided by modern CPUs to run the guest OS inside of the VM natively. 
This setup provides high performance virtualization. 
\nyx integrates with both \qemu and \kvm to take control of the VM, and to reset the state to a given snapshot. 
The fuzzer uses an agent component within the VM to control the fuzzing cycle: the agent indicates that the target is ready to receive an input and to create a snapshot. 
Then, the agent passes the input to the target and lastly the agent notifies the fuzzer that the test case was finished. 
To achieve this, the agent uses so-called hypercalls. 
Hypercalls are like syscalls but for VMs: they leave the VM context and pass the control to the hypervisor. \qemu then reacts to those events and creates or restores a snapshot.

\paragraph{\nyx Snapshots}
To take a snapshot, a copy of the physical memory and all device state is created. 
To revert back to a snapshot, all device state is overwritten by the old state.
Similarly, the VM's physical memory is overwritten by the original memory. 
Since the physical memory is often large (4GB and upwards), its prohibitively expensive to reset the whole memory.
To accelerate this process, both \nyx and \agamotto use a variety of optimizations: 
both fuzzers track which pages in the VM's memory have been altered since the start of the execution. 
This way they avoid overwriting the whole memory in favor of the (few) modified pages. 

Modern CPUs provide hardware acceleration features to support efficient tracking of the set of modified pages, the CPU tracks when a page is dirtied during execution. 
Once a certain amount of pages have been dirtied (typically up to 512 pages), the CPU exits the VM context and informs the hypervisor of the pages that were affected. 
The hypervisor, in this case \kvm, maintains a bitmap of pages that were written to. 
Both \nyx and \agamotto use this bitmap to selectively reset the VM's memory. 
However, \nyx's extension to \kvm also maintains a stack of pages that need to be reset. 
This allows \nyx to avoid searching the bitmap for pages to be reset after the execution. 
For some reason, \kvm uses 1 byte in the bitmap for each page in the physical memory. 
As a consequence, for a 4GB VM, \nyx's stack of dirty pages saves approximately 1MB of memory bandwidth per test case over \kvm's approach. 
Additionally, \nyx implements a custom reset mechanism for the state of emulated devices that is much faster than \qemu's native device serialization/deserialization routine.

%% file: sections/design.tex
\section{Design}
In the following, we describe the design of \tool and the rationale behind each choice we made when designing and building it. First, we give a short threat model that details the attack scenarios and surfaces that we concern ourselves with. Then, we present the architecture of \tool (see Figure~\ref{fig:arch} for a high-level overview).

\subsection{Threat Model}

Any network interface is usually a clear security boundary. 
As such, we mostly target various socket style interfaces (\eg TCP, UDP, and Unix domain sockets). 
However, in some context (such as sandboxes), this boundary sometimes also includes shared memory. 
We assume that the attacker has full control over all data that is being sent to the corresponding interfaces. 
With real networks, this is obvious: an attacker is usually able to send arbitrary data to any TCP/UDP server or client. 
In the case of Unix domain sockets and shared memory, we assume that the attacker has gained full control of the sandboxed process and attempts to exploit the higher privileged process controlling the sandbox.
As such, the attacker is also able to send arbitrary data via Unix domain sockets or shared memory.

\subsection{Towards Efficient Network Fuzzing}

Our approach uses \emph{hypervisor-based snapshot fuzzing}. Hence, the target is running in a customized VM.
At the same time, the fuzzer is running outside of the VM to ensure that the fuzzer has full control over the environment. 
Additionally, this method allows to take a snapshot of the target. 
During fuzzing, the snapshot is used to quickly reset the whole virtual machine back to a pristine state after each individual test case. 
This is already an excellent base to fuzz complex targets that communicate via network interfaces: the snapshot ensures that all state of network connections inside the VM is correctly reset between test cases. 
Even common, complex patterns such as forking a new process for each incoming connection, writing incoming data to a file system, or even a database in another process, are correctly handled. 

\pafig{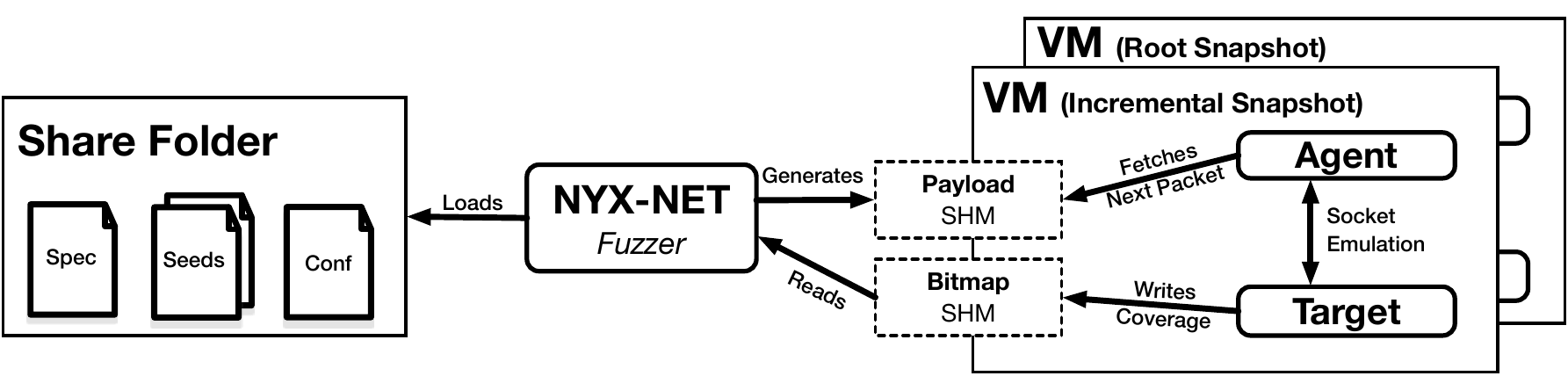}{fig:arch}{High-level overview of \tool's architecture. The fuzzer maintains two snapshots of the same VM, in which the agent component hooks the relevant connection of the target}

However, several distinct challenges remain: first, creating new network connections inside of the VM is still a slow process, usually involving dozens of context switches. 
This severely limits the fuzzing throughput. 
Second, many network protocols tend to be fundamentally slow: many messages need to be exchanged to reach ``interesting'' states. 
Often, a complex handshake has to be performed before data interacting with the actual application logic is exchanged. 
Last, the message formats are often complex and a precise understanding of each field or value involved is hard to obtain. 
As such, the purely generative approach used in \nyx is hard to use for network fuzzing.
Writing a specification that is precise enough to model all corner cases is cumbersome and would often take significant manual effort.
On the other hand, it is usually easy to obtain some traces of the communication with the target. 
To address these issues and to make fuzzing networking servers efficient and effective, we introduce the following techniques.

\subsection{High Performance Networking}
To ensure high performance networking and handle the complexity of shared memory based IPC mechanisms, we emulate significant fractions of the relevant functionality. 
To this end, we implement a library that adds a variety of hooks into existing \libc networking functionality. 
This library is injected into the target (e.g., via \texttt{LD\_PRELOAD} or by compiling it directly into the target) and intercepts all relevant calls. 
During startup of the target program, the hooks track various relevant interactions with the OS:
we track which file descriptors are part of the external attack surface and various metadata associated with them. 
When the fuzzing starts, it runs the \nyx bytecode VM to generate data reaching the target process on each hooked function call. 
Functions such as \texttt{read()} or \texttt{recv()} on target file descriptors consume data from packets encoded in the test case. 
More complex APIs such as \texttt{epoll()} are emulated to indicate which file descriptor (fd) is ready to receive data (\eg which fds are receiving packets next according to the bytecode). 
This library also ensures that packets are consumed correctly across multiple processes. 
Synchronizing the state of the bytecode stream is relevant if multiple processes are sharing file descriptors/sockets. 
For example, forking network servers will usually inherit a recently opened socket from the main process. 
Similarly, complex IPC protocols often contain the ability to share new file descriptors across existing socket connections.

By emulating network functionality, we gain the following advantages: first, we can precisely identify which data is attacker-controlled and inject our own data at the right places. Stemming from the same features, we can automatically infer the right place to create the initial snapshot. \tool automatically places the first snapshot after starting the process and directly before the first byte of input data is passed to the target. Second, and maybe even more importantly, we can often run a whole test case without hitting the slow operating system paths handling real network data. 
We can faithfully emulate network behaviors such as packets being received in discrete chunks. While TCP is a stream based protocol, at a first glance it often looks like TCP reads ``the same packets'' that were written by a single call to \texttt{send()}. While fundamentally broken, a frightening amount of servers assume that a single call to \texttt{recv()} will never return data from more than one ``packet'' (\eg corresponding \texttt{send()} call in the client). The same ability is also needed to properly emulate UDP connections, where packet boundaries are indeed semantic information.  

\subsection{Fuzzing with Incremental Snapshots} 
After ensuring network traffic is emulated with high precision and performance, we still face the challenge that in many cases, protocols contain long sequences of messages (and hence complex states). 
For example, \firefox's IPC traffic consists of hundreds and thousands of packets containing many kilobytes of relevant data. 
Even with fast network emulations, the number of such test cases that can be executed per second is strictly limited by the time needed to parse and consume these longs sequences. 
As an example, assume that we have a stream of 120 packets that we want to fuzz. 
Further, assume that we are currently fuzzing only the last 20 packets. 
Each test case will execute the same initial 100 packets over and over again. 
To overcome this hurdle for fuzzing performance, we add the ability to use incremental snapshots: 
\nyx starts each new execution from the root snapshot that represents a clean state. \tool adds the ability to quickly create and remove secondary snapshots after executing fragments of the input. This can be used to shave off common prefixes from the execution by taking an incremental snapshot after the common prefix was executed.
\pafig{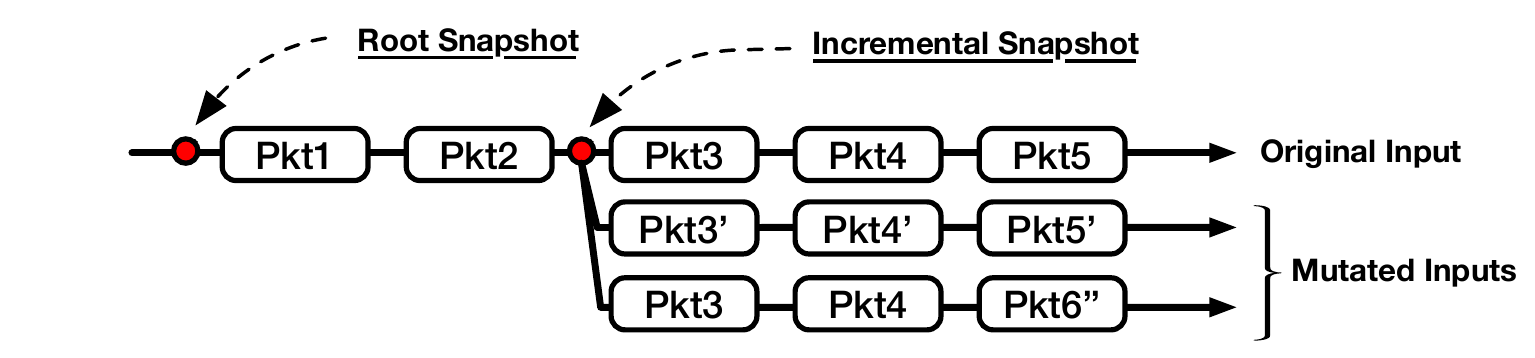}{fig:snapshots}{Using incremental snapshots to run mutated tests while skipping the common prefix consisting of packets one to three.}

For example, \tool starts from the root snapshot and executes the first 100 packets of a 120 packet sequence. 
It then take a secondary ``incremental'' snapshot that represents the VM state after executing the first 100 packets. 
Now, \tool runs a handful (tens to hundreds) of test cases that mutate only the last 20 packets. 
After each test case, we reset the VM state to the incremental snapshot taken after already processing the initial 100 packets. This process is visualized both in Figure~\ref{fig:smbc} and Figure~\ref{fig:snapshots}.
This saves the time required to handle those packets.
As soon as \tool wants to schedule another input, the incremental snapshot is discarded, and \tool returns to the root snapshot for the next input.

In some cases (especially slow targets) we have seen test throughput increase by more than 10x.
We took great care to ensure that the process of creating this secondary snapshot takes very little time. 
While this obviously introduced some engineering effort, it allows for great simplicity in other aspects of our design: we only ever keep one additional snapshot around. 
Creating incremental snapshots is so cheap that storing them would waste space and time. 
By recreating incremental snapshots on demand, we also avoid more complex structures such as trees of incremental snapshots building on top of each other~\cite{agamotto}. 
Each time a new input is scheduled for fuzzing, we randomly decide whether to use incremental snapshots for this input (depending on the inputs performance and number of packets).
Then, we pick a random packet in the input and create a snapshot after sending the given packet. 
Finally, we fuzz the remaining packets for a number of times before discarding the snapshot. 
In our experiments, we have seen that even for short state sequences reusing the snapshot as little as 50 times yields significant performance increases. 

\paragraph{Snapshot Scheduling}
A snapshot placement policy determines how the fuzzer selects the point at which it takes an incremental snapshot. 
To place snapshots, the fuzzer introduces a special ``snapshot'' opcode that can be injected at arbitrary positions in the input bytecode. The VM then creates an incremental snapshot when this opcode is executed.
\tool utilizes three strategies for snapshot scheduling.

\begin{description}
\item [\tool-none] As a baseline, running \tool without incremental snapshots is equivalent to a policy that always selects the root snapshot. 
\end{description}
Taking snapshots towards the end of the input sequence allows the fuzzer to better exploit its incremental snapshot capabilities; in some cases though, taking snapshots earlier enables the fuzzer to backtrack and possibly explore branching paths in the sequence. 
To this end, we implemented two snapshot placement policies to explore this trade-off. The parameters were empirically determined via small-scale studies and for sequences smaller than four packets, both policies select the root snapshot. 
\begin{description}
\item [\tool-balanced] On inputs with more than four packets, the balanced policy chooses the root snapshot in $4\%$ of the cases. Otherwise it selects a random index in the whole ($50\%$), or only in the second half ($50\%$). 

\item[\tool-aggressive] This policy cycles all available indices for snapshots. The first time an input is scheduled, it creates the snapshot at the end of the input. Each time no new inputs have been found by fuzzing this snapshot for $50$ iterations, we place the snapshot one packet earlier. When \tool-aggressive reaches the smallest index, it starts again from the end of the input. 
\end{description}

\subsection{Generating Complex Inputs}
As mentioned earlier, \nyx allows to express interactive protocols as input languages for fuzz targets by specifying opcodes for each possible interaction with the target. This features makes it easy to adapt \nyx to network fuzzing.
Unfortunately, the tool does not support to load existing network traces as seed files. 
This poses a serious restriction for network fuzzing. 
For \nyx, the data exchanged with a hypervisor or emulated hardware is usually structurally rather simple (individual pointers, bitfields, etc.). 
However, the data passed between clients and servers is often deeply nested and precisely modelling all aspects in \nyx's description mechanism is difficult. 
To ease the burden on the user, we introduce a way to convert raw packet dumps into \nyx bytecode inputs. 
This converter consists of a library that consumes \nyx's format specifications. It uses meta programming to create Python functions for each opcode. When we call those functions, the builder object logs each invocation. Finally, the builder outputs raw bytecode inputs to be used by our fuzzer. 

%% file: sections/implementation.tex
\section{Implementation Details}

To evaluate the performance of incremental snapshots and selective emulation, we implemented a prototype of \tool.
In the following section, we describe the implementation details of our fuzzer. 
We begin by describing the intricacies of emulating the network APIs used by real-world software, then we present the challenges and solutions of taking incremental snapshots, and lastly, we discuss turning \nyx's format specifications into a format that can be used to load complex existing network dumps as seeds. 

\subsection{Network Emulation}

To speed up network targets for fuzzing purposes and to inject our own fuzzing data, we emulate most network APIs. 
To be able to intercept calls to network APIs, we use an \texttt{LD\_PRELOAD} interceptor for common \libc functions. 
Obviously, we intercept common networking APIs such as \texttt{accept()}, \texttt{recv()}, etc. to track network sockets and the data sent to each socket.
However, we also emulate related APIs such as the \texttt{select/poll/epoll} interfaces to ensure compliant behavior. 
We also hook many APIs that operate on file descriptors in general, such as \texttt{dup()} and \texttt{close()} to keep track of aliasing file descriptors that are related to the targeted network connection.
For example, the \texttt{dup} family of operations is commonly used to pass file descriptors to child processes. 
Overall, our code hooks a total of 30 \libc functions and consists of roughly 2,000 lines of C code.

\subsection{Creating Incremental Snapshots}
\nyx only maintains a single root snapshot. 
Resetting the whole VM to this snapshot is very cheap: on small targets, \nyx is able to reset the VM about 12,000 times per second---about as fast as forking a similarly complex process once. 
While it is cheap to reset to the root snapshot, creating a root snapshot is expensive because it requires to copy the whole physical memory (often many gigabytes of data).

\tool extends \nyx's capabilities by introducing a second level snapshot that is much cheaper to create. 
This second level snapshot can be used to increase the performance on slow targets by skipping a whole prefix of each test case.
\tool makes taking an incremental snapshot about as cheap as resetting the snapshot once. 
As a consequence, we do not have to maintain complex data structures to store a set of snapshots like \agamotto. 
Instead, we simply recreate the snapshot for the current test case whenever needed. 

To obtain such a performance, we make use of similar facilities as resetting the original root snapshot. More specifically, we use \tool's ability to cheaply report the set of dirtied pages since the root snapshot was taken. The incremental snapshot can use this information to obtain a copy of all relevant memory. We also store another copy of \qemu's device state. 
To speed up resetting the VM to the incremental snapshot, we maintain a complete second mirror image of the VM's physical memory (see Figure~\ref{fig:arch}). 
However, to avoid creating an expensive copy of all the physical memory, we simply remap the existing root snapshot to a second location as Copy-On-Write pages. 
This way, the incremental snapshot itself looks like a complete root snapshot without incurring anywhere near the full memory cost.
As a consequence, we do not need to check whether to reset pages from the original root snapshot or the incremental one during the VM reset. 
To create the incremental snapshot, the pages that were dirtied by the execution since the root snapshot 
are overwritten with the content of the VM's physical memory. The Copy-On-Write mapping ensures that the original root snapshot remains unchanged.

Before creating another incremental snapshot, these pages are overwritten with the content of the root snapshot. 
Note that this means we accumulate real copies of pages already present in the root snapshot. 
In most cases, the executions affect the same memory. In these cases, reusing the existing copies avoids more expensive changes to the page tables.
However, in the worst case, this could lead to storing two identical copies of the root snapshot, causing twice the memory usage. 
To avoid this, we re-mirror the physical memory used in the incremental snapshot to a clean copy of the original root memory every 2,000 snapshots created.

To handle write accesses to emulated disks, \tool introduces a second caching layer to store dirtied sectors representing incremental snapshots. Like \nyx, we use a hashmap lookup to find sectors in the snapshot, otherwise we fall back to \nyx's root snapshot.

\subsection{Using Incremental Snapshots}
One of the core features of \nyx is to allow giving specifications for interactive targets. 
Each possible interaction is implemented as a small opcode that takes a set of arguments. 
When the fuzzer emits the opcode, the ``agent'' component performs whatever actions are requested. 
The opcode can also produce another set of values that may be used as arguments for future opcodes. 
Even though we are not using most of the features available in \nyx (e.g., affine types or even arguments/return values), this model is fundamentally a good fit for network fuzzing: 
the fuzzer is aware of the time dimension of each interaction. 
That is, the fuzzer knows about individual packets being sent and most importantly knows that packets that were not sent yet have also not affected the program state at all. 
This is crucial for incremental snapshots: 
we introduce a special ``snapshot'' opcode that the fuzzer injects at arbitrary positions in the input stream. 
When the agent encounters this packet, it requests a snapshot to be taken by a specific hypercall. 
Afterwards, the fuzzer continues fuzzing starting from the next packet only. 

\subsection{Creating Seed Files}

Since most hypercalls or MMIO accesses in emulated devices follow reasonably simple patterns, \nyx only supported specifying the interaction fully. 
No support for loading seed inputs exists in \nyx and the fuzzer needs to find all sequences of interactions on its own. 
This is not viable for network based fuzzing: protocols tend to be much more complex and it becomes prohibitively expensive to model them down to the last byte in \nyx's specification format. 
On the other hand, dumping network traffic is easy. 
As such, loading seed inputs adds tremendous value to fuzzing campaigns. 
To enable using PCAPs as seed inputs, we extended \nyx's specification engine with a Python library that allows to create inputs directly from Python code. 
The library consumes a specification and dynamically creates all function for each node. 
Each function logs the arguments and returns tracking objects that know which function call returned them. Later, calls that use those tracking objects as input can track where the values they use, were created. This way, the script builds a graph of function calls as well as their arguments and return values.
Finally, when calling \texttt{build()}, the graph is serialized into the flat bytecode that \nyx uses. 
An example seed file for the specification shown in Listing~\ref{lst:netmulti} can be seen in Listing~\ref{lst:netseed}.

We use this library in combination with \texttt{pyshark} to turn PCAP network dumps into seed files. 
To fragment TCP streams into logical packets, we use the same logic that \aflnet uses. 
While this is some protocol-specific code, the dissectors are usually very simple. 
For example, one of the more common packet boundary dissector uses the \texttt{CRLF} newline sequence to split the data stream into logical packets.

\begin{listing}[t]
\begin{minted}{py}
b = Builder(s)
con = b.connection()
b.packet(con, "HTTP/1.1 200 OK")
b.packet(con, "Content-Type: text/html")
\end{minted}
\vspace{-0.75em}
\caption{A manually created seed file for the multi-connection specification from Listing~\ref{lst:netmulti}.}
\label{lst:netseed}
\end{listing}

\subsection{Compile-Time Coverage}
\nyx supports only Intel PT to obtain coverage feedback. Yet, if available, compile-time instrumentation as introduced by \afl can be faster and more robust. To ease the use of \tool on platforms that do not support Intel PT and to improve performance on open-source targets, we enable compatibility with \afl's compile time instrumentations. The shared memory that contains the coverage bitmap is optionally exposed to the agent. The agent can then redirect \afl's coverage data to the shared memory that is used by QEMU. 

%% file: sections/evaluation.tex
\section{Evaluation}

\pafigwide{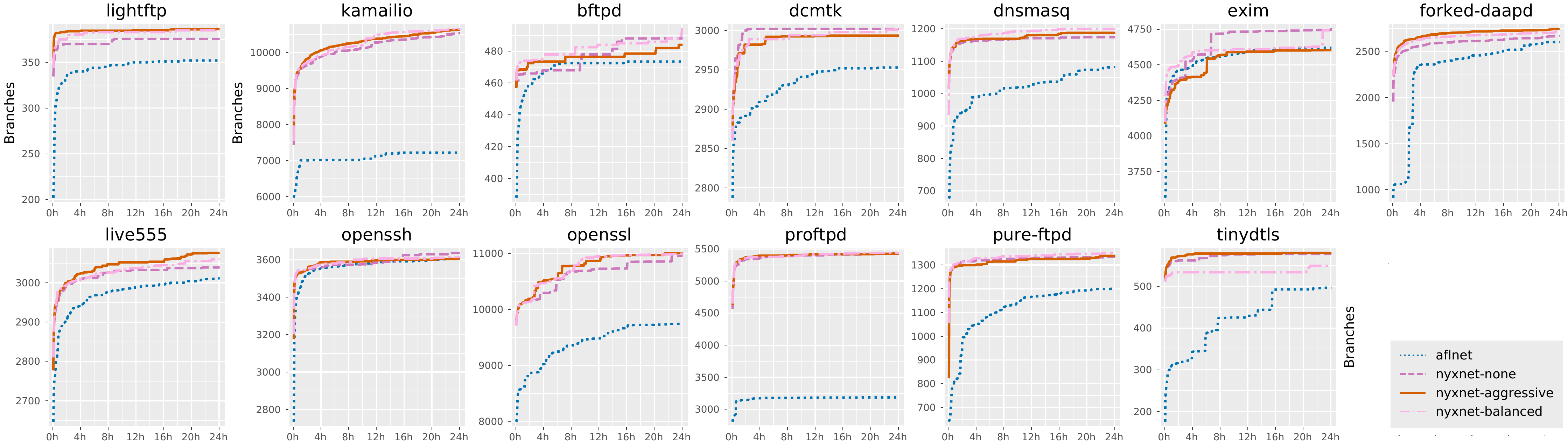}{fig:profuzzbench}{The median branch coverage across 10 experiments on all \profuzzbench targets. Note that we use the original plotting tools provided by \profuzzbench. This has two consequences: (i) the first measurement was taken after 10 seconds (sometimes hiding initial progress) and (ii) the y-axis is truncated to only show the coverage found after the seed files and hence does not start at 0.}
To evaluate the consequences of our design choices, we compare the prototype implementation of \tool both against baseline performance and other state-of-the-art network fuzzing tools.

As we will see, \tool outperforms the state-of-the-art in network service fuzzing on almost all targets. 
On the \profuzzbench benchmark for network fuzzing, \tool uncovers more coverage (usually between 10\% and 70\%).
Additionally, it usually reaches the same coverage between 10x to 100x, sometimes even 1000x faster (more details in the Appendix Table~\ref{tab:time-to-cov}).
In fact, on around half of the targets, \tool finds more coverage in the first five minutes than \aflnet in 24 hours. 
Additionally, \tool managed to find bugs in two targets of \profuzzbench that no other fuzzers is able to uncover. 
Lastly, we see that even for simple (in the case of \aflpp with \libpreeny) to moderately complex (in the case of \aflnet) targets, the approaches used by existing methods begin to fail in practice.
In contrast, \tool is not only able to handle all targets in the \profuzzbench suite, it even works for significantly more complex targets such as \firefox's IPC. 

\subsection{Evaluation Setup}
All experiments were performed on Intel Xeon Gold 6230 CPUs. 
Each machine had 52 physical cores and 192GB of memory as well as an SSD. 
When running experiments in parallel, each one was pinned to it's own physical core. 
We also disabled hyper-threading do reduce variance in performance. 
In experiments on \profuzzbench, we used \tool's ability to use \afl's compile-time instrumentation. 
This way, we can use the same target binary across all fuzzers.
Coverage experiments where repeated ten times and checked for statistical significance as recommended by Klees et al.~\cite{klees:fuzzEval}. 
We compared against \aflnet and \aflnwe in the most recent commits supported by \profuzzbench (0f51f9e and 6ba3a25). 
Likewise, we used a recent release of \aflpp (2dac4e7). 

\libpreeny contains two approaches used to turn network servers into targets suitable for fuzzing with \aflpp. 
The simple one only replaces sockets with \texttt{stdin} by hooking \texttt{accept()}. 
As this approach is unable to handle most real-world targets, \libpreeny also ships a more complex de-sockifying emulator. 
We found that it was able to handle more of the \profuzzbench targets. 
Hence, we chose to use the second, better performing emulation layer (\texttt{desock.c}).

\subsection{ProFuzzBench}
In the first experiment, we compare \tool against \aflnet and \aflpp in combination with \libpreeny's socket emulation layer.
Each individual fuzzing campaign was ran for 24h. We use the public \profuzzbench test suite. 
It contains a total of 13 different network services for various types of protocols (from FTP file transfer over VoIP to media streaming). Notably, \profuzzbench is published and maintained by the authors of \aflnet, and is used to showcase \aflnet's strength in fuzzing stateful network targets.
We used the coverage measurement and reporting features that are part of \profuzzbench.
A full set of final coverage results for all fuzzers is presented in Table~\ref{tab:pfb:cov} and coverage over time is also shown in Figure~\ref{fig:profuzzbench}. 
Note that on some targets \aflpp with \libpreeny is unable to even start the service. On most other targets, it makes some initial progress, but as coverage is only measured in five minute intervals, most or all coverage was found within the first five minutes and hence it seems that no coverage is found at all. Additionally, \aflnwe significantly under-performs compared to \aflnet. We therefore excluded both from Figure~\ref{fig:profuzzbench}. 
This results demonstrate how \libpreeny is far less powerful than our emulation layer. Similarly, \aflnet and \aflnet-no-state perform almost identical, and we excluded \aflnet-no-state from the plots. A complete figure containing plots of all fuzzers can be found in the Appendix (Figure~\ref{fig:pfb:complete}).
Overall, \tool is outperforming \aflnet on all but two targets that show no statistically significant difference.

\begin{table}[tb]

\scriptsize
\centering

\caption{Crashes found by each fuzzer in \profuzzbench. We excluded OOM crashes that were only due to the very narrow limits introduced by the docker setup of \profuzzbench. On dcmtk, \tool only finds crashes reliably if Asan is enabled (\cmark). This is due to the fact that in contrast to \aflnet, \tool does not build up memory corruption state until it crashes. With Asan, the crash is found within the first 10 seconds. Without Asan, \tool is able to find the bug in some runs, but not others depending on the initial memory layout. On pure-ftpd, \aflnet-no-state managed to trigger an OOM that was due to an internal limit and not the \profuzzbench limit (*). The targets that \aflpp+\libpreeny was unable to run are marked with n/a. We excluded targets where no fuzzer found anything of interest.}
\begin{threeparttable}
\begin{centering}
\begin{tabular}{r | c c c c c c}
\toprule
\multicolumn{1}{c}{} &\multicolumn{3}{c}{ \afl{}-based}&  \multicolumn{3}{c}{\tool} \\
     Target & \aflnet & \aflnwe & \aflpp & \textsc{none} & \textsc{balanced} & \textsc{aggressive} \\
     \hline
     dcmtk & \cmark & \cmark & n/a & (\cmark) & (\cmark) & \cmark \\
     dnsmasq & \cmark & \cmark & \cmark & \cmark & \cmark & \cmark \\
     exim  & - & - & n/a & \cmark &\cmark & \cmark \\
     live555 & \cmark & \cmark & n/a & \cmark & \cmark & \cmark \\
     proftpd & - & - & n/a & \cmark & \cmark & \cmark \\
     pure-ftpd & * & - & n/a & - & - & - \\
     tinydtls & \cmark & \cmark & n/a & \cmark & \cmark & \cmark \\
\bottomrule
\end{tabular}
\end{centering}
\end{threeparttable}
\label{tab:pfb:crashes}

\end{table}

We also investigated each tool's ability to find crashes in the targets contained in \profuzzbench. \aflnet, \aflnet-no-state and \aflnwe all find crashes in the exact same four targets. Similarly, \tool was able to crash the same four targets. Additionally, \tool also was able to crash two additional targets. A full list of the crashes uncovered can be found in Table~\ref{tab:pfb:crashes}.

\begin{centering}
\begin{table*}[tb]
    \small
    \centering
    \caption{Median branch coverage found by various fuzzers across 10 runs of 24h each, compared to \aflnet. The column for \aflnet displays the number of branches. All other columns show the changes compared to \aflnet. Changes that are statistically significant ($\rho<0.05$) according to a Mann-Whitney u-test are rendered bold.}
    \begin{threeparttable}

    \begin{tabular}{r|cccccccccccccc}
    \toprule
    \multicolumn{1}{c}{} &\multicolumn{4}{c}{ \afl{}-based}&  \multicolumn{3}{c}{\tool} \\
    & \aflnet & \aflnet-no-state & \aflnwe & \aflpp & \tool & \tool{}-balanced & \tool{}-aggressive \\
        \hline
bftpd & $473.5$ & $+1.3\%$ & $+1.6\%$ & n/a & $\mathbf{+3.1\%}$ & $\mathbf{+4.3\%}$ & $\mathbf{+2.2\%}$\\
dcmtk & $2953.0$ & $\mathbf{+0.7\%}$ & $\mathbf{+1.3\%}$ & n/a & $\mathbf{+1.7\%}$ & $\mathbf{+1.6\%}$ & $\mathbf{+1.4\%}$\\
dnsmasq & $1082.0$ & $-0.9\%$ & $\mathbf{-10.5\%}$ & $\mathbf{-37.2\%}$ & $\mathbf{+8.5\%}$ & $\mathbf{+10.9\%}$ & $\mathbf{+9.8\%}$\\
exim & $4620.5$ & $\mathbf{+1.5\%}$ & $\mathbf{-18.0\%}$ & n/a & $+2.9\%$ & $+2.7\%$ & $-0.4\%$\\
forked-daapd & $2604.5$ & $+0.1\%$ & $\mathbf{-23.4\%}$ & $\mathbf{-46.7\%}$ & $\mathbf{+2.5\%}$ & $\mathbf{+4.5\%}$ & $\mathbf{+5.5\%}$\\
kamailio & $7222.5$ & $-3.1\%$ & $\mathbf{-29.9\%}$ & n/a & $\mathbf{+45.9\%}$ & $\mathbf{+47.5\%}$ & $\mathbf{+47.2\%}$\\
lightftp & $352.0$ & $+0.4\%$ & $\mathbf{-53.4\%}$ & $\mathbf{-69.3\%}$ & $\mathbf{+6.7\%}$ & $\mathbf{+9.4\%}$ & $\mathbf{+9.8\%}$\\
live555 & $3011.5$ & $\mathbf{+1.3\%}$ & $\mathbf{+0.9\%}$ & n/a & $\mathbf{+0.9\%}$ & $\mathbf{+1.6\%}$ & $\mathbf{+2.1\%}$\\
openssh & $3609.0$ & $+0.3\%$ & $\mathbf{-2.0\%}$ & $\mathbf{-49.7\%}$ & $+0.8\%$ & $+0.2\%$ & $-0.1\%$\\
openssl & $9744.5$ & $-0.2\%$ & $\mathbf{-51.2\%}$ & $\mathbf{-18.8\%}$ & $\mathbf{+12.4\%}$ & $\mathbf{+13.0\%}$ & $\mathbf{+13.0\%}$\\
proftpd & $3186.5$ & $\mathbf{-0.9\%}$ & $+0.4\%$ & n/a & $\mathbf{+70.2\%}$ & $\mathbf{+70.7\%}$ & $\mathbf{+70.4\%}$\\
pure-ftpd & $1201.5$ & $+4.8\%$ & $+2.2\%$ & n/a & $\mathbf{+11.1\%}$ & $\mathbf{+12.4\%}$ & $\mathbf{+11.4\%}$\\
tinydtls & $497.0$ & $+3.9\%$ & $\mathbf{-38.8\%}$ & n/a & $\mathbf{+16.3\%}$ & $+10.6\%$ & $\mathbf{+16.8\%}$\\
    \bottomrule
    \end{tabular}
    \end{threeparttable}
    \label{tab:pfb:cov}
\end{table*}
\end{centering}

\begin{table*}[tb]
\small
\centering

\caption{Test throughput of various \afl based fuzzers and \tool configurations. Each entry shows the average \textbf{executions per second $\pm$ standard deviation} across our ten 24h runs. \tool-none is \tool without incremental snapshots. ``Aggressive'' and ``balanced'' denote the two different strategies each. It can be seen that aggressively using incremental snapshots drastically gives the highest test throughput in all cases. However, the biggest gains come from the root snapshot avoiding initialization all together.}

\begin{threeparttable}
\begin{centering}
\begin{tabular}{ r | r r r r | r r r  }
\toprule
\multicolumn{1}{c}{} &\multicolumn{4}{c}{ \afl{}-based}&  \multicolumn{3}{c}{\tool} \\
\hline
 Target & \aflnet & \aflnet-no-state  &  \aflnwe  &  \aflpp & \tool-none & \tool{}-balanced & \tool{}-aggressive \\
\midrule
bftpd & $4.2 \pm 1.9$ & $3.1 \pm 1.0$ & $3.5 \pm 1.8$ & - & $670.3 \pm 42.4$ & $1027.2 \pm 53.8$ & $\mathbf{1250.1} \pm 88.7$ \\
dcmtk & $33.8 \pm 2.3$ & $34.1 \pm 3.2$ & $38.4 \pm 1.0$ & - & $1716.7 \pm 127.3$ & $1673.9 \pm 246.5$ & $\mathbf{1782.5} \pm 96.3$ \\
dnsmasq & $3.3 \pm 1.6$ & $3.6 \pm 1.2$ & $1.6 \pm 2.0$ & $4.5 \pm 0.0$ & $2732.7 \pm 167.1$ & $2583.3 \pm 135.8$ & $\mathbf{2749.1} \pm 142.2$ \\
exim & $4.8 \pm 2.2$ & $4.4 \pm 2.5$ & $17.8 \pm 7.9$ & $69.3 \pm 0.2$ & $\mathbf{312.9} \pm 116.7$ & $307.9 \pm 92.1$ & $299.9 \pm 76.7$ \\
forked-daapd & $0.4 \pm 0.0$ & $0.4 \pm 0.0$ & $0.5 \pm 0.0$ & $1.2 \pm 0.0$ & $13.0 \pm 2.5$ & $13.5 \pm 1.7$ & $\mathbf{13.6} \pm 2.2$ \\
kamailio & $4.1 \pm 0.4$ & $4.3 \pm 0.2$ & $4.9 \pm 0.0$ & - & $274.8 \pm 19.4$ & $352.1 \pm 26.5$ & $\mathbf{624.6} \pm 56.6$ \\
lightftp & $6.1 \pm 1.5$ & $5.6 \pm 0.9$ & $13.0 \pm 6.6$ & $14.4 \pm 0.1$ & $1557.1 \pm 352.1$ & $1760.3 \pm 306.5$ & $\mathbf{2040.9} \pm 264.0$ \\
live555 & $7.7 \pm 2.6$ & $8.8 \pm 2.0$ & $25.9 \pm 8.8$ & - & $63.1 \pm 5.8$ & $84.2 \pm 11.5$ & $\mathbf{105.4} \pm 7.7$ \\
openssh & $23.6 \pm 8.3$ & $8.1 \pm 3.7$ & $29.3 \pm 1.0$ & $126.9 \pm 3.6$ & $136.4 \pm 3.3$ & $215.4 \pm 3.3$ & $\mathbf{830.2} \pm 33.2$ \\
openssl & $0.3 \pm 0.1$ & $0.8 \pm 1.6$ & $16.0 \pm 0.3$ & $16.8 \pm 0.6$ & $454.0 \pm 15.0$ & $\mathbf{467.1} \pm 21.6$ & $462.3 \pm 17.0$ \\
proftpd & $2.6 \pm 1.2$ & $1.7 \pm 0.7$ & $2.9 \pm 1.0$ & - & $332.7 \pm 37.1$ & $452.9 \pm 73.0$ & $\mathbf{518.1} \pm 149.8$ \\
pure-ftpd & $6.3 \pm 3.5$ & $5.8 \pm 1.3$ & $5.3 \pm 2.2$ & - & $849.8 \pm 56.9$ & $1450.8 \pm 61.5$ & $\mathbf{1806.1} \pm 120.2$ \\
tinydtls & $2.2 \pm 0.5$ & $2.2 \pm 0.2$ & $12.5 \pm 0.3$ & - & $1011.2 \pm 358.0$ & $942.3 \pm 283.5$ & $\mathbf{1228.0} \pm 315.7$ \\

 \bottomrule
\end{tabular}

\end{centering}
\end{threeparttable}
\label{tab:eval:execs}
\end{table*}

\subsection{Incremental Snapshots}

The targets that are part of \profuzzbench are configured in a way that \afl and its derivatives such as \aflnet and \aflnwe perform reasonably well. To this end, very short seeds with only a handful (\eg usually less than five) of packets where chosen by the authors. In such a scenario, high performance emulation and snapshot fuzzing make up most of the impact. While incremental snapshots still increase the throughput, they can not add their full potential. To evaluate the impact of incremental snapshot, we hence picked a more complex target with longer runs. Specifically, we demonstrate how incremental snapshots greatly increase the fuzzing throughput when fuzzing the game \emph{Super Mario Bros.} also used to showcase other fuzzing tools~\cite{ijon}. 

\paragraph{Super Mario} 
We recreate the Super Mario experiment presented in \ijon and demonstrate how \tool's incremental snapshots lead to 10x-30x increases over \ijon in time to solve. On all levels, \ijon was the slowest fuzzer. \tool-None added a modest 4x average speedup (standard deviation 2.4x, min/max: 1x/9.4x). \tool-Balanced managed to achieve an 5.8x average speedup (standard deviation: 3x, min/max: 1.6x/12.7x), while \tool-Aggressive found solutions on average 11x faster (standard deviation: 6.8x, min/max: 1.8, 29.8x). The data of all levels can be found in the Appendix (Figure~\ref{tab:eval:smbc}).
In fact, when fuzzing some of the simple levels on 52 cores in parallel, \tool is able to find a solution faster than a flawless player optimizing for speed (commonly known as ``speedrun'') is able to play the level even once. 
This unlikely feature is made possible by a combination of factors: 
most importantly, as can be seen in Figure~\ref{fig:smbc}, \tool's incremental snapshot allowed the fuzzer to focus only on the difficult part of the current execution by using incremental snapshots right in front of the difficult jump, leading to solve the level 10x -- 20x faster than \ijon.
Additionally, \ijon's experiment setup is skipping rendering and removes the framerate limit of 60 FPS. 

Lastly, we parallelize fuzzing to 52 cores. All these speedups together allow us to perform tens of thousands of test cases per second. 
As a consequence, \tool is able to solve the first level in less than the 26 seconds wall-clock time needed to speedrun the level at normal framerates. 

The original \ijon paper mentioned that \ijon was occasionally able to use wall jumps to escape from pits.
However, \tool actually was able to exploit this ability to much greater results: 
\tool is routinely able to solve a level (2-1) by exploiting a wall jump glitch. \ijon was unable to find this glitch and the authors of \ijon believed 2-1 might be impossible to solve.
\tool seems to be able to trigger this glitch somewhat regularly (it was found in two out of three of our configurations). 

\paragraph{Scalability}
It is important to be able to scale to many cores for fuzzing purposes.  Naively parallelizing the fuzzer like \agamotto or \nyx will consume prohibitive amounts of memory (e.g., many 100s or even 1,000s of GBs). We share the root snapshots between different instances. As a consequence, in our experiments, 80 instances of \tool only require about 2x the memory of a single instance. 

\paragraph{Snapshot Overhead}
To better understand the performance impact of incremental snapshots, we also perform detailed experiments evaluating the performance overhead introduced by our approach. To this end, we used three different policies of \tool for most experiments: \emph{None} (only a root snapshot is used), \emph{Balanced} (we are rather conservative about snapshots), and \emph{Aggressive} (almost every execution is using snapshots, and we are mostly placing the snapshot close to the end of the input).
This allows us to explore the impact of using incremental snapshots. Our experiments on \profuzzbench (seen in Table~\ref{tab:eval:execs}) show that while snapshots are an additional cost, aggressive snapshot produces the highest execution throughput on all targets. Even the balanced strategy still usually increases throughput. While it also reduces throughput in some cases, the difference is usually smaller than the variance between the different runs.
As mentioned before, \profuzzbench mostly consists of short sequences of inputs. 
As discussed earlier, when using incremental snapshots on \emph{Super Mario}, which has longer message sequences, more aggressive snapshots significantly improve the time to solve a target.

\pafig{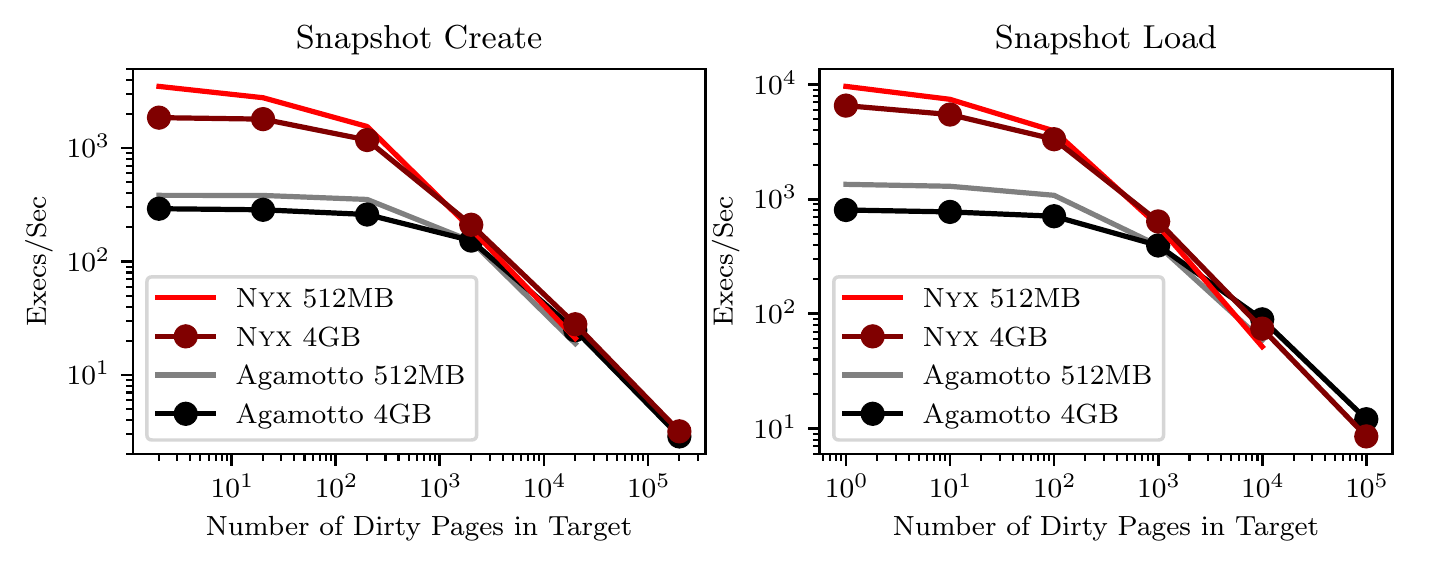}{fig:reloads}{Measuring the throughput of creating/loading incremental snapshots with $n$ dirty pages on VMs with 512MB and 4GB memory respectively.}

\paragraph{\agamotto}
Lastly, we compare our implementation of incremental snapshots against \agamotto, another recent fuzzer that was developed to speed up \syzkaller with incremental snapshots. 
We used both \agamotto's implementation and ours to create and restore incremental snapshots using the base VM image from our network experiments. 
We varied the number of dirtied pages and measured the time needed both for creating as well as resetting an incremental snapshot. 
Note that in contrast to the other experiments, this experiment was performed on a Intel Core i7-6700HQ CPU @ 2.60GHz as the larger servers were blocked by more computationally expensive experiments.
For creating snapshots, $n$ pages were dirtied, a temporary snapshot was created, and $n$ pages where dirtied again. Then, the old root snapshot was restored. This experiment was repeated 1,000 times each and the average times were measured. Note that the 500 MB VM was unable to dirty $10^5$ pages as not enough memory could be allocated. The results are show in Figure~\ref{fig:reloads}. 
A few notable results can be observed: first of all, \tool is almost an order of magnitude faster than \agamotto in the relevant range of dirtied pages. 
As expected, usually creating/restoring snapshots on smaller VMs is slightly faster. 
Surprisingly, for both \agamotto and \tool, restoring large numbers of dirty pages on the 512MB VM is slower than on the 4GB VM. 
This is due to the fact that allocating a significant fraction of the whole available memory is much more work than allocating the same number of pages if there is plenty of memory. 
Also, for large numbers of dirty pages, \agamotto becomes marginally faster than \tool. 
This is due to the fact that when the number of dirty pages approaches the number of pages available, the size of our dirty stack approaches (and eventually even exceeds) the size of the bitmap used. 
It should be noted that in contrast to \tool, \agamotto maintains a tree of snapshots. 
After one gigabyte is used to store snapshots, \agamotto begins to discard old snapshots causing it to slow down. 
This state is usually quickly reached during fuzzing (as well as in this experiment). 
We also performed offline experiments where we aborted before \agamotto would start using its LRU policy to evict snapshots. 
This increases \agamotto's performance, as no cleanups are performed. 
However, the performance was still behind \tool's throughput for typical workloads. 

Overall, \tool is significantly faster across the relevant part of the spectrum of snapshot sizes.
This observation is related to multiple factors. 
First, \tool uses a simpler mechanism: while \agamotto constructs trees of prefixed snapshots, \tool only uses a single snapshot. 
Second, \tool is not iterating the whole bitmap that tracks dirty pages, while \agamotto has to walk the whole bitmap of all pages present in the physical memory of the VM. 
Last, \tool also uses faster emulated device resets, reducing the fixed cost of resetting devices.

\subsection{Case Study: MySQL Client}

After we evaluated \tool on various server components, we now present a high-level view of using \tool for testing clients. For this case study, we fuzzed MySQL's client software that is used to connect to and administrate MySQL databases. 
Running \tool requires five steps: (i) obtain the target binary, (ii) pick or create a protocol specification, (iii) obtain seed inputs (optionally), (iv) bundles all required data, and (v) finally run the fuzzer.

\begin{enumerate}
\item  In a first step, we obtain a binary of MySQL client to fuzz by compiling the software with \afl's compiler.

\item Next, we need to chose or create a format specification. As we do not want to spend the time to learn about this protocol, we simply pick the generic default specification that assumes raw packets. 

\item To gather seed inputs in step three, we use Wireshark to obtain a set of PCAPs.
As the capture was taken locally, TCP packets directly correspond to logical packets in the protocol. 
Hence, we use the generic script to split the PCAP into individual packets used as seed.

\item The fourth step is to bundle a share folder that contains all relevant data. 
We use the packer script that copies the target, all of its dependencies, and the seeds into the share folder. 
It also parses the specification and auto-generates the LD\_PRELOAD library that is used as agent component during fuzzing. 

\item In the last step, we run the fuzzer by passing the path to the share folder. The fuzzer automatically loads the VM image, which runs a script that downloads the share directory and runs the target. 
\end{enumerate}
Performing these steps yields an out-of-bound read on the current version of the client (as shipped by Ubuntu) after a few minutes of fuzzing on 52 cores. We are currently in the process of reporting this issue.

\subsection{Case Study: Lighttpd}
We also used \tool on Lighttpd's development branch and found a memory corruption issue where a negative amount of memory could be allocated under specific circumstances. We reported the issue and the bug was fixed before being merged into master.

\subsection{Case Study: Firefox}

To demonstrate the versatility of \tool, we also fuzzed the IPC interface used by \firefox to separate high-risk, sandboxed content processes from the main process that contains all critical data. The fuzzing team at \firefox recently specifically asked for this kind of fuzzing in a public blog post~\cite{mozilla-ipc}. Luckily, \tool matches their needs very closely. It should be noted that the IPC interface is much more complex than the usual network services. It combines many different kinds of communication patters, from sockets, over shared memory to custom actor implementations used by JavaScript code to communicate between processes. Nonetheless, \tool is able to fuzz \firefox IPC with only some changes to the agent component (the LD\_PRELOAD library). 
Specifically, \firefox uses dozens of processes and threads and approximately a hundred sockets---many of which are needed at the same time. We extended the agent to find the relevant sockets and to allow the agent to talk to multiple connections at the same time.
While fuzzing \firefox, we found three bugs and the Firefox team found two additional security issues while evaluating and integrating \tool into their workflow. It should also be noted that we only fuzzed a very small subset of the available packets due to our limited understanding of \firefox's IPC mechanisms. After seeing the impact of \tool, the fuzzing team at Mozilla is currently planning to integrate \tool into their testing pipeline.
We were awarded a \$20.000 bug bounty for enabling fuzzing of the IPC interface of Firefox, mainly because our approach solves a long-standing problem at Mozilla.

\subsection{Handling of New Vulnerabilities}

We worked closely with a security engineer at Mozilla to understand and mitigate the security impact of the bugs found with \tool. While our three bugs where only null pointer dereferences (which are still regarded as high severity), the additional two bugs found by Mozilla were exploitable.
During the evaluation, we also found one crash in MySQL's client that affects the version of MySQL currently shipped in Ubuntu. We have not yet had the time to full triage this bug, but will responsibly disclose it to the maintainers after we fully triaged the issue (and before publication of this paper). Lastly, when fuzzing Lighttpd, we also uncovered an integer underflow in malloc that was fixed before it was shipped.
Additionally, \tool managed to find two new bugs in targets from the \profuzzbench suite. While no other fuzzer in our evaluation found these bugs, it seems like these bugs have been fixed in the latest release. 

%% file: sections/related_work.tex
\section{Related Work}

Following the publication of \afl~\cite{afl}, its impact soon caused a wave of additional research. Almost every design choice was investigated: \afl's input mutation algorithm where extended upon~\cite{aflsmart,nautilus,redqueen,codealchemist,zest,weizz} as was its ability to trigger and identify bugs~\cite{parmesan,hotfuzz,uaf-fuzz,memlock,fuzzsan,hotfuzz,slf}. 
To improve the strength of \afl's semi-random mutations, many researchers proposed to combine fuzzing with more elaborate program analysis techniques such as taint tracking~\cite{angora,vuzzer} and symbolic or concolic execution ~\cite{drillerfuzzer,dynamicinteger,dart,tfuzz,sage,qsym,taintscope,grammarwhite,digfuzz,dowsing,pangolin}.  

Similarly, as the fast coverage guidance is one of the defining features of \afl, it was heavily scrutinized. Additional feedback mechanism where invented and tested for improvements \cite{wangnot,retrowrite,collafl,instrim,steelix,untracer}. 
The last big component of \afl, after mutating and obtaining coverage feedback, is picking which input is fuzzed next. Like the other two components, input scheduling has been investigated thoroughly  ~\cite{ijon,entropic,scheduling,optimizing_seed_selection,program_adaptiv,aflfast,aflgo}. 
For a more in-depth overview of recent developments in fuzzing, please refer to  Man\`es et al.'s SoK paper~\cite{art-science-of-fuzzing}.
To make feedback fuzzing more applicable in various scenarios, the harnessing was improved. Fuzzers such as \syzkaller or \kafl~\cite{krace,triforce,kafl,syzkaller,usbfuzz,agamotto} adapted \afl's fuzzing model to kernel fuzzing. Fuzzers like \vdf and \nyx even target hypervisors~\cite{vdf,hypercube,nyx}. Hypervisor-based fuzzing is also commonly used to fuzz firmware~\cite{frankenstein,basesafe,firmafl,halucinator}
Snapshots were also used previously to speed up fuzzing. Besides \nyx, Falk proposed to use hypervisor-based snapshot fuzzing~\cite{cmilk}. Similarly, snapshots were used to improve fuzzing of Android apps~\cite{timetraveling}. \agamotto even uses incremental snapshots to accelerate kernel-level fuzzing. 

%% file: sections/conclusion.tex
\section{Conclusion}

In this paper, we present \tool, an approach to fuzz complex network services with high performance and fidelity. 
We believe that snapshot-based network fuzzing makes fuzzing significantly easier to use: the user does not have to ensure the absence of artifacts caused by vestigial state from earlier executions. 
At the same time, our app can also clearly outperform state-of-the-art approaches based on sending data via real network interfaces---often by orders of magnitude. We also found \tool very easy to use: using \nyx's support for binary-only fuzzing, we can directly take targets from Ubuntu's repositories and fuzz test them. Nonetheless, our network emulation is still not 100\% accurate in more complex scenarios (e.g., when multiple connection are needed at the same time). As such, we still needed to perform some changes to the agent when fuzzing \firefox's IPC. A more complete emulation would make \tool even easier to use, we leave this engineering challenge as future work.

%% file: sections/appendix.tex
\newpage
\onecolumn

\section*{Appendix}
\appendix

\

\section{Super Mario Experiment}

\begin{table}[h]
    \centering
    \caption{Times (HH:MM:SS) that various fuzzers need to solve Super Mario levels (median of three). 
    $\frac{n}{3}$ indicates that only $n$ out of the three runs solved the level. 
    The best configuration also shows the speedup over raw \ijon.}
\resizebox{0.49\textwidth}{!}{%
\begin{tabular}{ c | cc |cc cc cc }

\toprule
Level & \multicolumn{2}{c}{\ijon}  &  \multicolumn{2}{c}{\tool-none} & \multicolumn{2}{c}{\tool-balanced} & \multicolumn{2}{c}{\tool-aggressive}  \\
\midrule

1-1 & 00:46:35 \phantom{$\frac{1}{3}$} &  & 00:17:53 \phantom{$\frac{1}{3}$} &  & 00:07:36 \phantom{$\frac{1}{3}$} &  & 00:04:21 \phantom{$\frac{1}{3}$} & \textbf{(10.7x)}\\
1-2 & 02:52:36 \phantom{$\frac{1}{3}$} &  & 00:45:57 \phantom{$\frac{1}{3}$} &  & 00:38:33 \phantom{$\frac{1}{3}$} &  & 00:10:08 \phantom{$\frac{1}{3}$} & \textbf{(17.0x)}\\
1-3 & 00:39:31 \phantom{$\frac{1}{3}$} &  & 00:07:08 \phantom{$\frac{1}{3}$} &  & 00:07:40 \phantom{$\frac{1}{3}$} &  & 00:04:41 \phantom{$\frac{1}{3}$} & \textbf{(8.4x)}\\
1-4 & 00:29:00 \phantom{$\frac{1}{3}$} &  & 00:03:27 \phantom{$\frac{1}{3}$} &  & 00:09:59 \phantom{$\frac{1}{3}$} &  & 00:02:00 \phantom{$\frac{1}{3}$} & \textbf{(14.5x)}\\
2-1 &        -       &  & 11:21:19 $\frac{1}{3}$ &  & 03:12:49 $\frac{1}{3}$ &  &        -       & \\
2-2 & 01:02:30 \phantom{$\frac{1}{3}$} &  & 00:46:54 \phantom{$\frac{1}{3}$} &  & 00:33:16 \phantom{$\frac{1}{3}$} &  & 00:23:14 \phantom{$\frac{1}{3}$} & \textbf{(2.7x)}\\
2-3 & 01:47:13 \phantom{$\frac{1}{3}$} &  & 00:34:42 \phantom{$\frac{1}{3}$} &  & 00:08:27 \phantom{$\frac{1}{3}$} &  & 00:03:36 \phantom{$\frac{1}{3}$} & \textbf{(29.8x)}\\
2-4 & 01:22:52 \phantom{$\frac{1}{3}$} &  & 00:09:31 \phantom{$\frac{1}{3}$} &  & 00:07:48 \phantom{$\frac{1}{3}$} &  & 00:04:16 \phantom{$\frac{1}{3}$} & \textbf{(19.4x)}\\
3-1 & 03:33:49 \phantom{$\frac{1}{3}$} &  & 00:35:42 \phantom{$\frac{1}{3}$} &  & 00:24:14 \phantom{$\frac{1}{3}$} &  & 00:09:45 \phantom{$\frac{1}{3}$} & \textbf{(21.9x)}\\
3-2 & 01:14:07 \phantom{$\frac{1}{3}$} &  & 00:18:13 \phantom{$\frac{1}{3}$} &  & 00:08:23 \phantom{$\frac{1}{3}$} &  & 00:05:33 \phantom{$\frac{1}{3}$} & \textbf{(13.4x)}\\
3-3 & 00:15:59 \phantom{$\frac{1}{3}$} &  & 00:05:04 \phantom{$\frac{1}{3}$} & \textbf{(3.2x)} & 00:05:18 \phantom{$\frac{1}{3}$} &  & 00:05:25 \phantom{$\frac{1}{3}$} & \\
3-4 & 00:31:54 \phantom{$\frac{1}{3}$} &  & 00:10:56 \phantom{$\frac{1}{3}$} &  & 00:10:50 \phantom{$\frac{1}{3}$} &  & 00:02:50 \phantom{$\frac{1}{3}$} & \textbf{(11.3x)}\\
4-1 & 01:18:24 \phantom{$\frac{1}{3}$} &  & 00:08:19 \phantom{$\frac{1}{3}$} &  & 00:09:35 \phantom{$\frac{1}{3}$} &  & 00:04:09 \phantom{$\frac{1}{3}$} & \textbf{(18.9x)}\\
4-2 & 00:22:42 \phantom{$\frac{1}{3}$} &  & 00:22:23 \phantom{$\frac{1}{3}$} &  & 00:13:50 \phantom{$\frac{1}{3}$} &  & 00:12:19 \phantom{$\frac{1}{3}$} & \textbf{(1.8x)}\\
4-3 & 00:31:36 \phantom{$\frac{1}{3}$} &  & 00:09:14 \phantom{$\frac{1}{3}$} &  & 00:06:37 \phantom{$\frac{1}{3}$} & \textbf{(4.8x)} & 00:07:39 \phantom{$\frac{1}{3}$} & \\
4-4 & 00:10:33 \phantom{$\frac{1}{3}$} &  & 00:03:55 \phantom{$\frac{1}{3}$} &  & 00:02:54 \phantom{$\frac{1}{3}$} &  & 00:01:42 \phantom{$\frac{1}{3}$} & \textbf{(6.2x)}\\
5-1 & 00:55:11 \phantom{$\frac{1}{3}$} &  & 00:25:54 \phantom{$\frac{1}{3}$} &  & 00:12:12 \phantom{$\frac{1}{3}$} &  & 00:07:00 \phantom{$\frac{1}{3}$} & \textbf{(7.9x)}\\
5-2 & 01:30:17 \phantom{$\frac{1}{3}$} &  & 00:13:31 \phantom{$\frac{1}{3}$} &  & 00:10:19 \phantom{$\frac{1}{3}$} &  & 00:08:38 \phantom{$\frac{1}{3}$} & \textbf{(10.5x)}\\
5-3 & 00:23:29 \phantom{$\frac{1}{3}$} &  & 00:14:24 \phantom{$\frac{1}{3}$} &  & 00:05:29 \phantom{$\frac{1}{3}$} &  & 00:05:00 \phantom{$\frac{1}{3}$} & \textbf{(4.7x)}\\
5-4 & 00:12:31 \phantom{$\frac{1}{3}$} &  & 00:11:17 \phantom{$\frac{1}{3}$} &  & 00:04:57 \phantom{$\frac{1}{3}$} &  & 00:04:17 \phantom{$\frac{1}{3}$} & \textbf{(2.9x)}\\
6-1 & 00:40:28 \phantom{$\frac{1}{3}$} &  & 00:13:42 \phantom{$\frac{1}{3}$} &  & 00:12:40 \phantom{$\frac{1}{3}$} &  & 00:05:11 \phantom{$\frac{1}{3}$} & \textbf{(7.8x)}\\
6-2 & 05:09:52 $\frac{2}{3}$ &  & 02:29:49 \phantom{$\frac{1}{3}$} &  & 01:18:50 \phantom{$\frac{1}{3}$} &  & 00:31:54 \phantom{$\frac{1}{3}$} & \textbf{(9.7x)}\\
6-3 & 00:27:01 \phantom{$\frac{1}{3}$} &  & 00:03:20 \phantom{$\frac{1}{3}$} & \textbf{(8.1x)} & 00:03:57 \phantom{$\frac{1}{3}$} &  & 00:05:22 \phantom{$\frac{1}{3}$} & \\
6-4 & 00:29:28 \phantom{$\frac{1}{3}$} &  & 00:05:33 \phantom{$\frac{1}{3}$} &  & 00:03:36 \phantom{$\frac{1}{3}$} &  & 00:01:35 \phantom{$\frac{1}{3}$} & \textbf{(18.6x)}\\
7-1 & 01:39:41 \phantom{$\frac{1}{3}$} &  & 00:14:21 \phantom{$\frac{1}{3}$} &  & 00:10:09 \phantom{$\frac{1}{3}$} &  & 00:06:29 \phantom{$\frac{1}{3}$} & \textbf{(15.4x)}\\
7-2 & 01:18:41 \phantom{$\frac{1}{3}$} &  & 00:41:15 \phantom{$\frac{1}{3}$} &  & 00:35:25 \phantom{$\frac{1}{3}$} &  & 00:24:16 \phantom{$\frac{1}{3}$} & \textbf{(3.2x)}\\
7-3 & 01:49:48 \phantom{$\frac{1}{3}$} &  & 00:45:11 \phantom{$\frac{1}{3}$} &  & 00:17:21 \phantom{$\frac{1}{3}$} &  & 00:05:08 \phantom{$\frac{1}{3}$} & \textbf{(21.4x)}\\
7-4 & 00:52:48 \phantom{$\frac{1}{3}$} &  & 00:09:41 \phantom{$\frac{1}{3}$} &  & 00:04:38 \phantom{$\frac{1}{3}$} & \textbf{(11.4x)} & 00:07:08 \phantom{$\frac{1}{3}$} & \\
8-1 & 11:10:34 $\frac{2}{3}$ &  & 03:57:19 \phantom{$\frac{1}{3}$} &  & 01:24:57 \phantom{$\frac{1}{3}$} &  & 00:49:27 \phantom{$\frac{1}{3}$} & \textbf{(13.6x)}\\
8-2 & 05:04:01 \phantom{$\frac{1}{3}$} &  & 03:43:08 \phantom{$\frac{1}{3}$} &  & 00:45:00 \phantom{$\frac{1}{3}$} &  & 00:28:48 \phantom{$\frac{1}{3}$} & \textbf{(10.6x)}\\
8-3 & 00:27:01 \phantom{$\frac{1}{3}$} &  & 00:09:50 \phantom{$\frac{1}{3}$} &  & 00:10:04 \phantom{$\frac{1}{3}$} &  & 00:03:34 \phantom{$\frac{1}{3}$} & \textbf{(7.6x)}\\

\bottomrule
\end{tabular}}
\label{tab:eval:smbc}
\end{table}

\section{Time to Equal Coverage}

\begin{table}[h]    
    \centering

    \caption{This table shows at what time \aflnet found its final coverage and how much faster \tool was able to obtain the same coverage. Note that in some cases \tool is able to find more coverage in a minute than \aflnet does in 24h, and the time resolution of our coverage measurements is limited to one minute. As \tool never exceeded \aflnet's coverage on exim, the values are omitted. These numbers should be taken with a grain of salt, as even small increases in coverage can easily translate to significant improvements in terms of "Time to Equal Coverage" due to the logarithmic nature of new coverage found.}
    \scriptsize \centering
  \begin{tabular}{lc|ccc}
    Target & \aflnet time to Final Coverage & \tool & \tool-balanced & \tool-aggressive\\
    \hline
   bftpd          & 15:32:00& 1x & 24x & 6x\\
 dcmtk          & 21:34:00& 51x & 41x & 28x\\
 dnsmasq        & 23:32:00& 235x & 201x & 282x\\
 exim           & 19:31:00& 2x & 1x & - \\
 forked-daapd   & 23:19:00& 2x & 8x & 14x\\
 kamailio       & 15:10:00& 910x & 910x & 910x\\
 lightftp       & 18:38:00& 139x & 372x & 1118x\\
 live555        & 23:30:00& 5x & 5x & 7x\\
 openssh        & 23:47:00& 1x & 1x & -\\
 openssl        & 23:54:00& 717x & 717x & 717x\\
 proftpd        & 23:42:00& 1422x & 1422x & 1422x\\
 pure-ftpd      & 23:27:00& 234x & 469x & 234x\\
 tinydtls       & 22:58:00& 1378x & 1378x & 1378x\\
        \bottomrule
  \end{tabular}
    \label{tab:time-to-cov}
\end{table}

\newpage

\section{Coverage Plots}
\begin{figure}[h!]
  \centering
      \includegraphics[width=\linewidth]{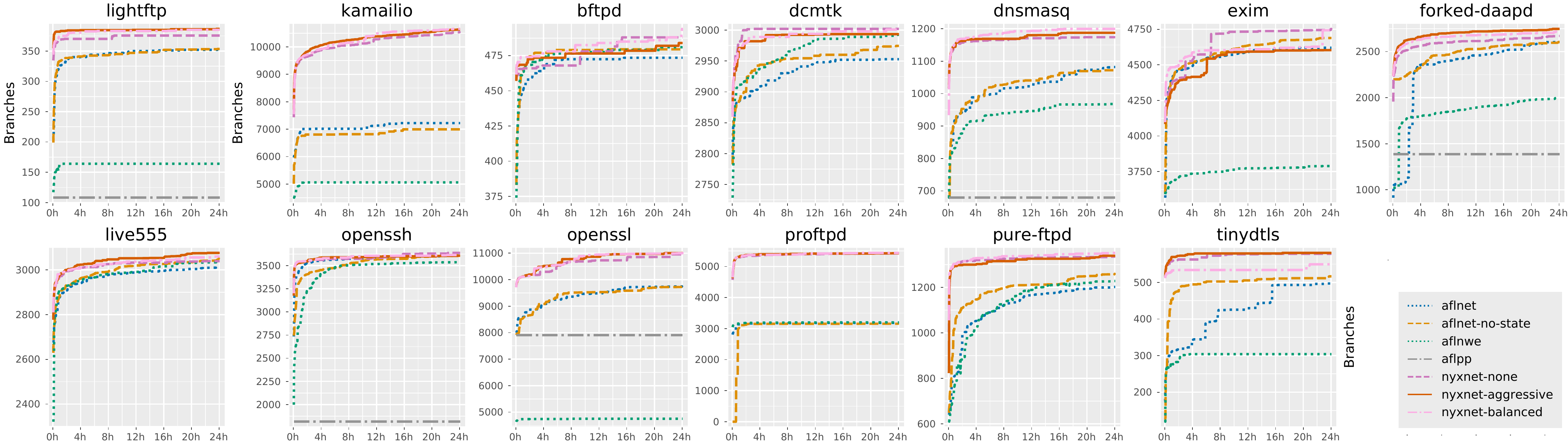}
  \caption{The median branch coverage across 10 experiments of all fuzzers on all \profuzzbench targets.}
  \label{fig:pfb:complete}
\end{figure}